\pdfoutput=1
\documentclass[aip,	apl,reprint,graphicx,amsmath,amsfonts,amssymb,,floatfix]{revtex4-1}

\usepackage[utf8]{inputenc} %
\usepackage[english]{babel}
\usepackage[babel]{microtype} %
\usepackage{bm} %
\usepackage{xparse}
\usepackage{mathtools}
\usepackage{mleftright}
\usepackage{dsfont}
\usepackage{textcomp} %
\usepackage{siunitx}
\usepackage{array} %
\usepackage{tabularx} %
\usepackage{booktabs} %
\usepackage{chemformula}
\usepackage{tikz}
\usetikzlibrary{calc, backgrounds, arrows.meta, positioning, fadings, 3d, angles, shapes.geometric, decorations.pathreplacing, decorations.pathmorphing, decorations.text, fit}
\usepackage{hyperref}
\usepackage{nameref}
\usepackage[resetlabels]{multibib}

\newcites{SI}{REFERENCES}

\usepackage{xcolor}
\definecolor{uablue}{RGB}{0,68,102}
\colorlet{uablue100}{uablue}
\colorlet{uablue75} {uablue!75!white}
\colorlet{uablue50} {uablue!50!white}
\colorlet{uablue25} {uablue!25!white}
\colorlet{uablue10} {uablue!10!white}
\colorlet{uablue5}  {uablue!5!white}

\definecolor{uared}{RGB}{136,17,51}
\colorlet{uared100}{uared}
\colorlet{uared75} {uared!75!white}
\colorlet{uared50} {uared!50!white}
\colorlet{uared25} {uared!25!white}
\colorlet{uared10} {uared!10!white}
\colorlet{uared5}  {uared!5!white}

\definecolor{ualightblue}{RGB}{51,153,204}
\colorlet{ualightblue100}{ualightblue}
\colorlet{ualightblue75} {ualightblue!75!white}
\colorlet{ualightblue50} {ualightblue!50!white}
\colorlet{ualightblue25} {ualightblue!25!white}
\colorlet{ualightblue10} {ualightblue!10!white}
\colorlet{ualightblue5}  {ualightblue!5!white}

\definecolor{uagold}{RGB}{221,153,17}
\colorlet{uagold100}{uagold}
\colorlet{uagold75} {uagold!75!white}
\colorlet{uagold50} {uagold!50!white}
\colorlet{uagold25} {uagold!25!white}
\colorlet{uagold10} {uagold!10!white}
\colorlet{uagold5}  {uagold!5!white}

\definecolor{uayellow}{RGB}{170,170,0}
\colorlet{uayellow100}{uayellow}
\colorlet{uayellow75} {uayellow!75!white}
\colorlet{uayellow50} {uayellow!50!white}
\colorlet{uayellow25} {uayellow!25!white}
\colorlet{uayellow10} {uayellow!10!white}
\colorlet{uayellow5}  {uayellow!5!white}

\definecolor{darkgrey}{RGB}{60,60,59}
\definecolor{lightgrey}{RGB}{146,148,151}

\colorlet{dcolor}{uablue} %
\colorlet{dcolor100}{dcolor}
\colorlet{dcolor75} {dcolor!75!white}
\colorlet{dcolor50} {dcolor!50!white}
\colorlet{dcolor25} {dcolor!25!white}
\colorlet{dcolor10} {dcolor!10!white}
\colorlet{dcolor5}  {dcolor!5!white}

\colorlet{lcolor}{uared} %
\colorlet{lcolor100}{lcolor}
\colorlet{lcolor75} {lcolor!75!white}
\colorlet{lcolor50} {lcolor!50!white}
\colorlet{lcolor25} {lcolor!25!white}
\colorlet{lcolor10} {lcolor!10!white}
\colorlet{lcolor5}  {lcolor!5!white}

\colorlet{emphcolor}{uared}
\colorlet{emphcolor100}{emphcolor}
\colorlet{emphcolor75} {emphcolor!75!white}
\colorlet{emphcolor50} {emphcolor!50!white}
\colorlet{emphcolor25} {emphcolor!25!white}
\colorlet{emphcolor10} {emphcolor!10!white}
\colorlet{emphcolor5}  {emphcolor!5!white}

\colorlet{addcolor}{uagold} %
\colorlet{addcolor100}{addcolor}
\colorlet{addcolor75} {addcolor!75!white}
\colorlet{addcolor50} {addcolor!50!white}
\colorlet{addcolor25} {addcolor!25!white}
\colorlet{addcolor10} {addcolor!10!white}
\colorlet{addcolor5}  {addcolor!5!white}
\makeatletter
\newcommand\RedeclareMathOperator{%
  \@ifstar{\def\rmo@s{m}\rmo@redeclare}{\def\rmo@s{o}\rmo@redeclare}%
}
\newcommand\rmo@redeclare[2]{%
  \begingroup \escapechar\m@ne\xdef\@gtempa{{\string#1}}\endgroup
  \expandafter\@ifundefined\@gtempa
     {\@latex@error{\noexpand#1undefined}\@ehc}%
     \relax
  \expandafter\rmo@declmathop\rmo@s{#1}{#2}}
\newcommand\rmo@declmathop[3]{%
  \DeclareRobustCommand{#2}{\qopname\newmcodes@#1{#3}}%
}
\@onlypreamble\RedeclareMathOperator
\makeatother
\makeatletter
\DeclareFontFamily{OMX}{MnSymbolE}{}
\DeclareSymbolFont{MnLargeSymbols}{OMX}{MnSymbolE}{m}{n}
\SetSymbolFont{MnLargeSymbols}{bold}{OMX}{MnSymbolE}{b}{n}
\DeclareFontShape{OMX}{MnSymbolE}{m}{n}{
	<-6>  MnSymbolE5
	<6-7>  MnSymbolE6
	<7-8>  MnSymbolE7
	<8-9>  MnSymbolE8
	<9-10> MnSymbolE9
	<10-12> MnSymbolE10
	<12->   MnSymbolE12
}{}
\DeclareFontShape{OMX}{MnSymbolE}{b}{n}{
	<-6>  MnSymbolE-Bold5
	<6-7>  MnSymbolE-Bold6
	<7-8>  MnSymbolE-Bold7
	<8-9>  MnSymbolE-Bold8
	<9-10> MnSymbolE-Bold9
	<10-12> MnSymbolE-Bold10
	<12->   MnSymbolE-Bold12
}{}

\let\llangle\@undefined
\let\rrangle\@undefined
\DeclareMathDelimiter{\llangle}{\mathopen}%
{MnLargeSymbols}{'164}{MnLargeSymbols}{'164}
\DeclareMathDelimiter{\rrangle}{\mathclose}%
{MnLargeSymbols}{'171}{MnLargeSymbols}{'171}
\makeatother
\newcommand{\opm}{
  \mathbin{
    \mathchoice
      {\buildcirclepm{\displaystyle     }{0.14ex}{0.95}{0.05ex}{.7}}
      {\buildcirclepm{\textstyle        }{0.14ex}{0.95}{0.05ex}{.7}}
      {\buildcirclepm{\scriptstyle      }{0.13ex}{0.955}{0.04ex}{.55}}
      {\buildcirclepm{\scriptscriptstyle}{0.08ex}{0.95}{0.03ex}{.45}}
  }
}
\newcommand\buildcirclepm[5]{%
  \begin{tikzpicture}[baseline=(X.base), inner sep=-#5, outer sep=-.65]
    \node[draw,circle,line width=#4] (X)  {\footnotesize\raisebox{#2}{\scalebox{#3}{$#1\pm$}}};
  \end{tikzpicture}%
}
	\DeclarePairedDelimiter{\opbrack}{(}{)}
	\NewDocumentCommand{\oplbrack}{s O{} m}{%
		\IfBooleanTF{#1}
		{\mleft(#3\mright.\kern-\nulldelimiterspace}
		{\mathopen{#2(} #3 \mathclose{}}%
	}
	\NewDocumentCommand{\oprbrack}{s O{} m}{%
		\IfBooleanTF{#1}
		{\kern-\nulldelimiterspace\mleft.#3\mright)}
		{\mathopen{} #3 \mathclose{#2)}}%
	}
	
	\NewDocumentCommand{\opsqlbrack}{s O{} m}{%
		\IfBooleanTF{#1}
		{\mleft[#3\mright.\kern-\nulldelimiterspace}
		{\mathopen{#2[} #3 \mathclose{}}%
	}
	\NewDocumentCommand{\opsqrbrack}{s O{} m}{%
		\IfBooleanTF{#1}
		{\kern-\nulldelimiterspace\mleft.#3\mright]}
		{\mathopen{} #3 \mathclose{#2]}}%
	}

	\NewDocumentCommand{\bracketlr}{s O{} m m m m m}{%
		\IfBooleanTF{#1}
		{\left#4 #3 \right#5^{#6}_{#7}}
		{\mathinner{ \mathopen{#2#4} #3 \mathclose{#2#5}^{#6}_{#7} } }%
	}
	\NewDocumentCommand{\bracketl}{s O{} m m}{%
		\IfBooleanTF{#1}
		{ \left#4 #3 \right.\kern-\nulldelimiterspace }
		{ \mathinner{ \mathopen{#2#4} #3 \mathclose{} } }%
	}
	\NewDocumentCommand{\bracketr}{s O{} m m m m}{%
		\IfBooleanTF{#1}
		{ \kern-\nulldelimiterspace\left. #3 \right#4^{#5}_{#6} }
		{ \mathinner{ \mathopen{} #3 \mathclose{#2#4}^{#5}_{#6} } }%
	}

	\NewDocumentCommand{\nbrack}{s O{} m t^G{} t_G{} }{%
		\IfBooleanTF{#1}
		{ \bracketlr*{#3}{(}{)}{#5}{#7} }
		{ \bracketlr[#2]{#3}{(}{)}{#5}{#7} }
	}
	\NewDocumentCommand{\nlbrack}{s O{} m}{%
		\IfBooleanTF{#1}
		{ \bracketl*{#3}{(} }
		{ \bracketl[#2]{#3}{(} }%
	}
	\NewDocumentCommand{\nrbrack}{s O{} m t^G{} t_G{} }{%
		\IfBooleanTF{#1}
		{ \bracketr*{#3}{)}{#5}{#7} }
		{ \bracketr[#2]{#3}{)}{#5}{#7} }%
	}

	\NewDocumentCommand{\sqbrack}{s O{} m t^G{} t_G{} }{%
		\IfBooleanTF{#1}
		{ \bracketlr*{#3}{[}{]}{#5}{#7} }
		{ \bracketlr[#2]{#3}{[}{]}{#5}{#7} }
	}
	\NewDocumentCommand{\sqlbrack}{s O{} m}{%
		\IfBooleanTF{#1}
		{ \bracketl*{#3}{[} }
		{ \bracketl[#2]{#3}{[} }%
	}
	\NewDocumentCommand{\sqrbrack}{s O{} m t^G{} t_G{} }{%
		\IfBooleanTF{#1}
		{ \bracketr*{#3}{]}{#5}{#7} }
		{ \bracketr[#2]{#3}{]}{#5}{#7} }%
	}

	\NewDocumentCommand{\cbrack}{s O{} m t^G{} t_G{} }{%
		\IfBooleanTF{#1}
		{ \bracketlr*{#3}{\{}{\}}{#5}{#7} }
		{ \bracketlr[#2]{#3}{\{}{\}}{#5}{#7} }
	}
	\NewDocumentCommand{\clbrack}{s O{} m}{%
		\IfBooleanTF{#1}
		{ \bracketl*{#3}{\{} }
		{ \bracketl[#2]{#3}{\{} }%
	}
	\NewDocumentCommand{\crbrack}{s O{} m t^G{} t_G{} }{%
		\IfBooleanTF{#1}
		{ \bracketr*{#3}{\}}{#5}{#7} }
		{ \bracketr[#2]{#3}{\}}{#5}{#7} }%
	}

	\NewDocumentCommand{\abrack}{s O{} m t^G{} t_G{} }{%
		\IfBooleanTF{#1}
		{ \bracketlr*{#3}{\langle}{\rangle}{#5}{#7} }
		{ \bracketlr[#2]{#3}{\langle}{\rangle}{#5}{#7} }
	}
	\NewDocumentCommand{\albrack}{s O{} m}{%
		\IfBooleanTF{#1}
		{ \bracketl*{#3}{\langle} }
		{ \bracketl[#2]{#3}{\langle} }%
	}
	\NewDocumentCommand{\arbrack}{s O{} m t^G{} t_G{} }{%
		\IfBooleanTF{#1}
		{ \bracketr*{#3}{\rangle}{#5}{#7} }
		{ \bracketr[#2]{#3}{\rangle}{#5}{#7} }%
	}

	\DeclarePairedDelimiter{\abs}{\lvert}{\rvert}
	\DeclarePairedDelimiter{\norm}{\lVert}{\rVert}

	\NewDocumentCommand{\evalat}{s O{} m m}{%
	  \IfBooleanTF{#1}
	    {\kern-\nulldelimiterspace\left.#3\right|_{#4}}
	    {#3#2|_{#4}}%
	}

	\DeclarePairedDelimiterX{\closedinterval}[2]{[}{]}{#1,\,#2}
	\DeclarePairedDelimiterX{\openinterval}[2]{]}{[}{#1,\,#2}
	\DeclarePairedDelimiterX{\leftopeninterval}[2]{]}{]}{#1,\,#2}
	\DeclarePairedDelimiterX{\rightopeninterval}[2]{[}{[}{#1,\,#2}

	\DeclarePairedDelimiterX\Set[1]\{\}{%
		
		#1
	}

	\ExplSyntaxOn
	\NewDocumentCommand{\coord}{sO{}m}
	{
		\IfBooleanTF{#1}
		{\mleft(\coord_print:n {#3}\mright)}
		{\mathopen{#2(}\coord_print:n {#3}\mathclose{#2)}}
	}
	
	\seq_new:N \l_coord_list_seq
	\tl_new:N \l_coord_last_tl
	\cs_new_protected:Npn \coord_print:n #1
	{
		\seq_set_split:Nnn \l_coord_list_seq { , } { #1 } %
		\seq_pop_right:NN \l_coord_list_seq \l_coord_last_tl
		\seq_map_inline:Nn \l_coord_list_seq { ##1 , } %
		\tl_use:N \l_coord_last_tl
	}
	\ExplSyntaxOff

\newcommand{\infrac}[2]{{\begingroup #1 \endgroup /{#2}}}
\newcommand{\Eq}[1]{Eq.~(\ref{#1})}

\newcommand{\Fig}[1]{Fig.~\ref{#1}}
\def\etal.{et\penalty\hyphenpenalty\ al.}
\newcommand*{\andothers}{et\penalty\hyphenpenalty\ al.}

\renewcommand*{\vec}[1]{\bm{\mrm{#1}}}
\newcommand{\vk}{\vec{k}}

\newcommand{\mrm}[1]{\mathrm{#1}}
\newcommand*{\mathmrm}[1]{\relax\ifmmode\mrm{#1}\else{#1}\fi} %

\newcommand{\cosfn}{\cos\opbrack}
\newcommand{\sinfn}{\sin\opbrack}
\newcommand{\cothfn}{\coth\opbrack}
\newcommand{\cossq}[1]{\cos^2{#1}}
\newcommand{\sinsq}[1]{\sin^2{#1}}
\newcommand{\ii}{\mathrm{i}}
\newcommand{\ee}{\mathrm{e}}
\newcommand{\expof}[1]{\ee^{#1}}
\DeclareMathOperator*{\Realpart}{Re}
\DeclareMathOperator*{\Imaginarypart}{Im}
\newcommand{\Over}[1]{\frac{1}{#1}}
\newcommand{\inOver}[1]{\infrac{1}{#1}}
\newcommand{\dint}[4]{\int_{#1}^{#2}#3\,\mrm{d}#4}

\newcommand{\OP}[1]{\vphantom{\mrm{#1}}\smash[t]{\hat{\mrm{#1}}}}
\newcommand*{\OPv}[1]{\vphantom{\vec{\mrm{#1}}}\smash[t]{\hat{\vec{\mrm{#1}}}}}
\newcommand{\inv}[1]{#1^{-1}}
\DeclareMathOperator{\Tr}{Tr}
\newcommand{\OPA}{\OP{A}}
\newcommand{\OPB}{\OP{B}}
\newcommand{\OPO}{\OP{O}}
\newcommand{\OPS}[2]{{\OP{S}_{#2}^{#1}}}
\newcommand*{\OPvS}{\OPv{S}}
\newcommand{\lvec}[1]{\vec{r}_{#1}} %

\newcommand{\inpitwo}{\infrac{\pi}{2}}
\newcommand{\bcdot}{\mathbin{\bm{\cdot}}}
\newcommand*{\conjugate}[1]{\overline{#1}}
\newcommand*{\posvec}{\vec{r}} %
\newcommand*{\posdiffvec}{\vec{r}} %
\newcommand*{\recprimcellvol}{v_b} %
\newcommand*{\fbrillzone}{\mrm{BZ}} %
\newcommand*{\lconst}{a} %
\newcommand*{\unitvec}{\vec{e}} %
\newcommand*{\kB}{k_\mrm{B}} %
\newcommand*{\kBT}{k_\mrm{B}T} %
\newcommand*{\muB}{\mu_\mrm{B}} %
\newcommand*{\landeg}{g_{\text{e}}} %
\newcommand*{\Bmag}{B} %
\newcommand*{\Bangle}{\theta_{\Bmag}} %
\newcommand*{\vBmag}{\vec{\Bmag}} %
\newcommand*{\nBeff}{b_{\text{eff}}}
\newcommand*{\nBmag}{b} %

\newcommand*{\xcrys}{X} %
\newcommand*{\ycrys}{Y} %
\newcommand*{\zcrys}{Z} %
\newcommand*{\xmcs}{x} %
\newcommand*{\ymcs}{y} %
\newcommand*{\zmcs}{z} %
\newcommand*{\pmcs}{+} %
\newcommand*{\mmcs}{-} %
\newcommand*{\pmmcs}{\pm} %
\newcommand*{\exJ}{J} %
\newcommand*{\nexJ}{\eta} %
\newcommand*{\exJF}{\exJ} %
\newcommand*{\nexJF}{\nexJ} %
\newcommand*{\nanexJ}{\iota} %
\newcommand*{\nanexJF}{\nanexJ} %
\newcommand*{\anexJ}{I} %
\newcommand*{\anexJF}{\anexJ} %
\newcommand*{\texJ}{\exJF\opbrack{0}} %
\newcommand*{\tnanexJ}{\delta} %
\newcommand*{\Mag}{M} %
\newcommand*{\vMag}{\vec{\Mag}} %
\newcommand{\anis}{\Delta} %
\newcommand*{\srfunc}{\zeta} %
\newcommand*{\srint}{{\cal P}} %
\newcommand*{\Curie}{\mrm{C}} %
\newcommand*{\corr}{C} %
\newcommand*{\vcorr}{\vec{\corr}} %
\newcommand*{\exval}{c} %
\newcommand*{\tcorr}{\mathcal{\corr}} %
\newcommand*{\vtcorr}{\vec{\tcorr}} %
\DeclarePairedDelimiter{\corrf}{\langle}{\rangle} %
\DeclarePairedDelimiter{\expval}{\langle}{\rangle} %
\DeclarePairedDelimiterX{\commutator}[2]{[}{]}{#1,\,#2} %
\newcommand*{\OPH}{\OP{H}} %
\newcommand*{\OPHex}{\OP{H}_\text{ex}} %
\newcommand*{\OPHB}{\OP{H}_\text{B}} %
\newcommand*{\subA}{\mathmrm{A}} %
\newcommand*{\subB}{\mathmrm{B}} %
\newcommand*{\subGen}{\mathmrm{\circ}} %
\newcommand*{\subGenp}{\mathmrm{\diamond}} %
\newcommand*{\Even}{\mathmrm{E}} %
\newcommand*{\Odd}{\mathmrm{O}} %
\renewcommand*{\Re}{\mathmrm{R}} %
\renewcommand*{\Im}{\mathmrm{I}} %
\newcommand*{\Sat}{\mathmrm{S}} %

\newcommand{\sinda}{{\vec{p}}} %
\newcommand{\sindb}{{\vec{j}}}
\newcommand{\sindc}{{\vec{d}}}
\newcommand{\sindd}{{\vec{l}}}
\newcommand{\sindapos}{\posvec_{\sinda}} %
\newcommand{\sindbpos}{\posvec_{\sindb}}
\newcommand{\sinddpos}{\posvec_{\sindd}}
\newcommand{\hca}{a}

\newcommand{\hcfe}{f_{\Even}}
\newcommand{\hcfr}{f_{\Re}}
\newcommand{\hcfi}{f_{\Im}}
\newcommand{\hcfgen}{f_{\alpha}}
\newcommand{\hcge}{g_{\Even}}
\newcommand{\hcgr}{g_{\Re}}
\newcommand{\hcgi}{g_{\Im}}
\newcommand{\hcggen}{g_{\alpha}}
\newcommand*{\Green}{G} %
\newcommand*{\vGreen}{\vec{\Green}} %
\newcommand*{\tGreen}{\mathcal{\Green}} %
\newcommand*{\vtGreen}{\vec{\tGreen}} %
\DeclarePairedDelimiterX{\greenf}[2]{\llangle}{\rrangle}{#1;\,#2} %

\newcommand{\one}{\mathds{1}}

\newcommand{\vGamma}{\vec{\Gamma}}
\newcommand{\vOmega}{\vec{\Omega}}
\newcommand{\vPsi}{\vec{\Psi}}

\newcommand{\vL}{\vec{L}}
\newcommand{\vR}{\vec{R}}

\newcolumntype{Y}{ >{\raggedleft\arraybackslash$}X<{$}}
\newcolumntype{Z}{ >{\raggedright\arraybackslash$}X<{$}}
\begin{document}

\newcommand*{\ourtitle}{2D ferromagnetism at finite temperatures under quantum scrutiny}
\title{\ourtitle}

\newcommand*{\UAaffiliation}{%
	Department of Physics, University of Antwerp, Groenenborgerlaan 171, B-2020 Antwerp, Belgium%
}
\newcommand*{\UAnano}{%
	NANOlab Center of Excellence, University of Antwerp, Belgium%
}
\newcommand*{\IMECaffiliation}{%
	IMEC, Kapeldreef 75, B-3001 Leuven, Belgium%
}
\newcommand*{\KULaffiliation}{%
	ESAT, KU Leuven, Kasteelpark Arenberg 10, B-3001 Leuven, Belgium%
}

\author{Joren \surname{Vanherck}}
\affiliation{\UAaffiliation}
\affiliation{\IMECaffiliation}

\author{Cihan \surname{Bacaksiz}}
\affiliation{\UAaffiliation}
\affiliation{\UAnano}

\author{Bart \surname{Sor\'ee}}
\affiliation{\UAaffiliation}
\affiliation{\IMECaffiliation}
\affiliation{\KULaffiliation}

\author{Milorad V. \surname{Milo\v{s}evi\'c}}
\email{milorad.milosevic@uantwerpen.be}
\affiliation{\UAaffiliation}
\affiliation{\UAnano}

\author{Wim \surname{Magnus}}
\email{wim.magnus@uantwerpen.be}
\affiliation{\UAaffiliation}
\affiliation{\IMECaffiliation}

\date{\today}

\begin{abstract}
	Recent years have seen a tremendous rise of two-dimensional (2D) magnetic materials, several of which verified experimentally.
	However, most of the theoretical predictions to date rely on ab-initio methods, at zero temperature and fluctuations-free, while one certainly expects detrimental quantum fluctuations at finite temperatures.
	Here we present the solution of the quantum Heisenberg model for honeycomb/hexagonal lattices with anisotropic exchange interaction up to third nearest neighbors and in an applied field in arbitrary direction, that answers the question whether long-range magnetization can indeed survive in the ultrathin limit of materials, up to which temperature, and what the characteristic excitation (magnon) frequencies are, all essential to envisaged applications of magnetic 2D materials.
	We find that long-range magnetic order persists at finite temperature for materials with overall easy-axis anisotropy.
	We validate the calculations on the examples of monolayer \ch{CrI3}, \ch{CrBr3} and \ch{MnSe2}.
	Moreover, we provide an easy-to-use tool to calculate Curie temperatures of new 2D computational materials.
\end{abstract}

\maketitle %

Seminal experiments of Huang \andothers\cite{Huang17}\ on mechanically exfoliated \ch{CrI3} marked the advent of ferromagnetic two-dimensional (2D) materials in 2017.
Ever since, the related field of research has been at the forefront of theoretical and experimental investigations (for review, see Refs. \onlinecite{Gong19,Gibertini19}), with a plethora of materials and heterostructures found to exhibit exciting magnetic properties in their thinnest limit.

Over the last decade, the world has seen an enormous scientific investment in emergent phenomena of 2D materials, where magnetism is only the latest addition.
The interest in these van der Waals (vdW) materials---typically of honeycomb or hexagonal lattice types---is further driven by their technological promises.
Concerning 2D magnetism, vdW materials offer precise magnetostatic control, heterostructure engineering and uniform thickness, which can benefit magnetic memories (such as magnetic tunnel junctions (MTJs)\cite{Parkin04} or spin-transfer torque magnetorestitive random-access memory (STT-MRAM)\cite{Kent15}), spintronic devices\cite{Zutic04, Fert08}, spin-wave logic\cite{Chumak15,Csaba17,Fischer17}, even quantum computing\cite{Khitun01}.
More exotic states in honeycomb lattices, such as topological magnon insulators\cite{Owerre16}, generate ideas towards topological edge transport with dissipationless magnons\cite{Chen18}.
The importance and feasibility of these potential applications was recently further strengthened by evidences of 2D ferromagnetism above room temperature in monolayer \ch{VSe2}\cite{Bonilla18} and \ch{MnSe2}\cite{OHara18}.
Experimental and technological advances are further enabled by new imaging techniques, probing magnetization directions\cite{Kim19} and nanometre-scale spatial variations\cite{Anahory20,Mattiat20} in 2D materials.

Before such high technological promises come to real applications, a thorough theoretical understanding of long-range 2D ferromagnetism as a function of temperature is required, beyond the first-principles description of spin interactions and magnetocrystalline anisotropy from Density Functional Theory (DFT).
The underlying physics governing the appearance of finite magnetization in a 2D material despite the Mermin-Wagner theorem\cite{Mermin66}---strictly speaking only valid in the absence of both sufficient anisotropy and slow decaying long-range interactions\cite{Nakano95}---and the effect of the reduced dimensionality on the susceptibility are still not comprehended.
To date, most descriptions\cite{Zhuang15} of Honeycomb spin lattices were either classical\cite{Torelli18} (or even of Ising type), take into account only interactions between nearest neighbors (NNs), or bosonize the spins using the Holstein-Primakoff\cite{Holstein40} approximation that is only justified for high spin values or at the lowest temperatures\cite{Pershoguba18}.
For studies of more complex systems or device architectures, both quantum mechanical challenges and atomistic discretization need to be neglected altogether to apply micromagnetic simulations\cite{Kim10,Wang18,Mueller19}.

\begin{figure}[t]
	\includegraphics{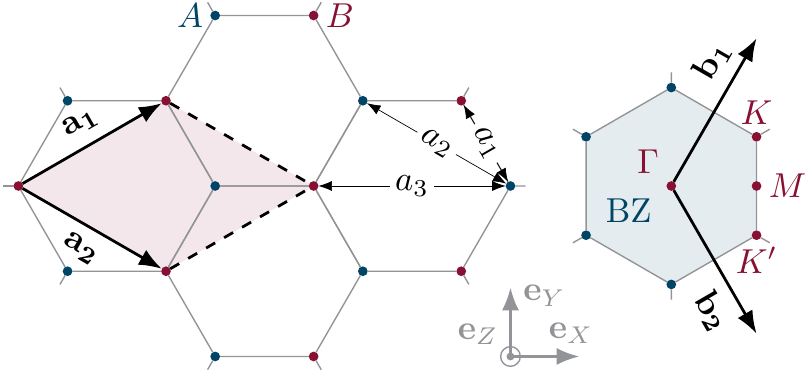}
	\caption{%
		The honeycomb lattice in direct space (left) and reciprocal space (right) as oriented in the crystallographic $\xcrys\ycrys$-plane.
		The honeycomb lattice is a hexagonal Bravais lattice with basis vectors $\vec{a_1}$ and $\vec{a_2}$, spanning the primitive cell (shaded in red), and a two-atom basis.
		These two sublattices are labeled \subA{} and \subB{}.
		The n\textsuperscript{th} NN distances $\lconst_{n}$ are indicated.
		In the reciprocal space, the reciprocal basis vectors $\vec{b_1}$ and $\vec{b_2}$, the first Brillouin zone $\fbrillzone$ (shaded in blue) and high-symmetry points $\Gamma$, $M$, $K$ and $K'$ are indicated.%
		}%
	\label{fig:honlattice}
\end{figure}

To change this dissatisfying picture in the 2D realm of magnetism, we here solve the quantum Heisenberg model for hexagonal and honeycomb lattices (see \Fig{fig:honlattice}), with anisotropic exchange interaction up to the third NN, and in applied field in an arbitrary direction.
Using microscopic parameters obtained from DFT, we go on to calculate the Curie temperature, excitation spectra, and full magnetization characteristics of selected recently discovered two-dimensional ferromagnets, validating the stability of their magnetization against temperature and quantum fluctuations, and their potential for magnonic and other applications.

We describe these materials as a lattice of effective spins $\OPvS_{\sindc}$ of magnitude $S$ interacting through a quantum Heisenberg Hamiltonian $\OPH= \OPHB + \OPHex$.
The interaction with an external magnetic field $\vBmag= \Bmag \nbrack{\unitvec_{\xcrys} \sin{\Bangle} + \unitvec_{\zcrys}\cos{\Bangle} }$, applied at an angle $\Bangle$ with respect to the out-of-plane direction $\unitvec_{\zcrys}$, is $\OPHB= -\landeg\muB \vBmag \bcdot \sum_{\sindc} \OPvS_{\sindc}$,
with land\'e g-factor $\landeg \approx 2$, Bohr magneton $\muB$ and $\hbar=1$.
It suffices to consider $0 \leqslant \Bangle \leqslant \inpitwo$ due to inversion symmetry.
The spins mutually interact through the exchange interaction
\begin{eqnarray}
	\OPHex 
	= 
		- \Over{2} \smash{\sum_{\sindc} \sum_{\sindd}} \exJ_{\sindd\sindc}
		\sqlbrack[\big]{
			\nbrack{
				1-\anis_{\sindd\sindc}
			} 
			\nbrack{
				\OPS{\xcrys}{\sindc}\OPS{\xcrys}{\sindd} 
				+ \OPS{\ycrys}{\sindc}\OPS{\ycrys}{\sindd}
			}
		} \nonumber \\ \sqrbrack[\big]{
			+\nbrack{
				1+\anis_{\sindd\sindc}
			}
			\OPS{\zcrys}{\sindc}\OPS{\zcrys}{\sindd}
		},
\end{eqnarray}
where spins at lattice sites $\sindc$ and $\sindd$ interact with exchange strength $\exJ_{\sindd\sindc}$ and anisotropy $\anis_{\sindd\sindc}$.
The interaction is ferromagnetic when $\exJ_{\sindd\sindc}>0$, in which case positive $\anis_{\sindd\sindc}$ favors the out-of-plane spin component $\OPS{\zcrys}{\sindd}$.
The exchange interaction and anisotropy are equally strong for n\textsuperscript{th} NNs: $\exJ_{\sindd\sindc}=\exJ_n$ and $\anis_{\sindd\sindc}=\anis_n$ when $\sindd$ and $\sindc$ indicate n\textsuperscript{th} NN lattice sites.
We consider the interactions up to third nearest neighbors, so $\exJ_n=0$ for $n>3$.

To effectively describe the above model, we formally define a general short-ranged interaction between lattice sites $\sindc$ and $\sindd$ at a distance $\norm{\posdiffvec_{\sindd\sindc}}$ as
\begin{equation}
	\srint_{\sindd\sindc}
	\coloneqq
		\sum_{n=1}^3 \srint_n \srfunc_{n, \sindd\sindc},
		\qquad
	\srfunc_{n, \sindd\sindc}
	\coloneqq 
		\delta\opbrack{\norm{\posdiffvec_{\sindd\sindc}}, \lconst_{n}},
\end{equation}
with $\srint_n$ and $\lconst_n$ the interaction strength and distance between n\textsuperscript{th} NNs respectively.
Two specific types of short-range interactions of the above form need to be distinguished: (i) the normalized exchange interaction $\nexJ_{\sindd\sindc} \coloneqq	\sum_{n=1}^3 \nexJ_n \srfunc_{n, \sindd\sindc}$,
with n\textsuperscript{th} NN normalized exchange strengths $\nexJ_n=\infrac{\exJ_n}{\texJ}$ and the total exchange strength $\texJ=3\nbrack{\exJ_1+2\exJ_2+\exJ_3}>0$ for an overall ferromagnetic interaction; and (ii)
the normalized anisotropy-weighted exchange interaction $\nanexJ_{\sindd\sindc}	\coloneqq \sum_{n=1}^3 \nanexJ_n \srfunc_{n, \sindd\sindc}$,
with $\nanexJ_n=\infrac{\anexJ_n}{\texJ}$ and $\anexJ_n = \anis_n \exJ_n$.
We define the total weighted anisotropy as $\tnanexJ \coloneqq \infrac{\anexJF\opbrack{0}}{\texJ} =	3\nexJ_1 \anis_1 + 6\nexJ_2 \anis_2 + 3\nexJ_3 \anis_3$.

The spatial Fourier transform of the n\textsuperscript{th} NN general short-ranged interaction for an atom on sublattice $\subGen\in\Set{\subA,\subB}$ is $\srint_{n, \subGen}\opbrack{\vk}= \srint_n \srfunc_{n, \subGen}\opbrack{\vk}$,
with
\begin{equation}
	\srfunc_{n, \subGen}\opbrack{\vk}
	=
		\sum_{\sindd} \srfunc_{n, \sindd\sindc} \expof{\ii \vk\bcdot\nbrack{\posvec_\sindd-\posvec_\sindc}}
		, \quad \sindc \in \subGen.
\end{equation}
Note that $\srfunc_{2}\opbrack{\vk}\coloneqq \srfunc_{2, \subGen}\opbrack{\vk}$ is independent of the reference sublattice, while $\srfunc_{m, \subB}\opbrack{\vk} = \conjugate{\srfunc_{m, \subA}\opbrack{\vk}}$ are complex valued for $m\in \Set{1,3}$, because they describe the interaction between different sublattices.
We have $\srfunc_{m, \Re}\opbrack{\vk}\coloneqq	\Realpart \srfunc_{m, \subA}\opbrack{\vk}$ as the real ($\Re$) and $\srfunc_{m, \Im}\opbrack{\vk}	\coloneqq \Imaginarypart \srfunc_{m,\subA}\opbrack{\vk}$ as the imaginary ($\Im$) parts of these interactions.
Finally, we define an even ($\Even$) and an odd ($\Odd$) part of the general short-ranged interaction as $\srint_{\Even}\opbrack{\vk}	\coloneqq \srint_n \srfunc_{2}\opbrack{\vk}$
and
\begin{align}
	\abs{\srint_{\Odd}\opbrack{\vk}}^{2}
	\coloneqq{}&
		\srint_{\Re}^2\opbrack{\vk} + \srint_{\Im}^2\opbrack{\vk},
		\qquad\text{where} \\
	\srint_{\Re}\opbrack{\vk}
	\coloneqq{}&
		\srint_1 \srfunc_{1,\Re}\opbrack{\vk} + \srint_3 \srfunc_{3,\Re}\opbrack{\vk} \\
	\srint_{\Im}\opbrack{\vk}
	\coloneqq{}&
		\srint_1 \srfunc_{1,\Im}\opbrack{\vk} + \srint_3 \srfunc_{3,\Im}\opbrack{\vk}.
\end{align}

Our solution method for this model is as follows (details in Supplementary Information (SI), Sec.~\ref{SI::sol}).
First, the original Hamiltonian $\OPH$, which is described in the crystallographic coordinate system $\Set{\unitvec_\xcrys,\unitvec_\ycrys,\unitvec_\zcrys}$, is expressed in terms of the magnetization coordinate system $\Set{\unitvec_\xmcs,\unitvec_\ymcs,\unitvec_\zmcs}$, with $\unitvec_\zmcs$ parallel to the yet unknown magnetization direction.
This is achieved most generally by rotation over an angle $\theta$ ($0\leqslant \theta \leqslant \inpitwo$) around the $\unitvec_\ycrys=\unitvec_\ymcs$ axis.
Next, we define the double-time temperature-dependent Green's functions\cite{Zubarev60,Callen63} $\Green^{\subGen,\alpha}_{\sinda,\sindb}\opbrack{\omega, \lambda}\coloneqq \greenf{\OPS{\alpha}{\sinda}}{\expof{ \lambda \OPS{\zmcs}{\sindb} } \OPS{\mmcs}{\sindb}	}$ with $\sinda \in \subGen$
for each $\alpha\in\Set{\pmcs, \mmcs, \zmcs}$, which obey the equation of motion 
\begin{equation}
	\omega \Green^{\subGen,\alpha}_{\sinda,\sindb} 
	= 
		\Over{2\pi} \expval{
			\commutator{\OPS{\alpha}{\sinda}}{\expof{ \lambda \OPS{\zmcs}{\sindb} } \OPS{\mmcs}{\sindb}} 
		} \delta_{\sinda\sindb}
		+ \greenf{ 
			\commutator{\OPS{\alpha}{\sinda}}{\OPH} 
		}{
			\expof{ \lambda \OPS{\zmcs}{\sindb} } \OPS{\mmcs}{\sindb} 
		}
\end{equation}
in frequency space.
Here we introduced real parameter $\lambda$ and $\OPS{\pmmcs}{\sinda}\coloneqq \OPS{\xmcs}{\sinda} \pm \ii\OPS{\ymcs}{\sinda}$.
Due to the two-atom basis, these equations of motion now constitute a coupled set of six instead of three\cite{Vanherck18} equations to solve.
The last term in the equation of motion contains higher-order Green's functions, which we decouple using the Tyablikov\cite{Tyablikov59} (RPA\cite{Englert60}) decoupling approximation, effectively neglecting second-order correlations between spins at different lattice sites.
The smart choice of the magnetization coordinate system ensures that terms containing $\expval{ \OPS{\pmmcs}{\sinda} }=0$ vanish.
After exploiting the translational symmetry using a spatial Fourier transformation, this yields a singular set of six algebraic equations.
The singularity is removed by using the regularity condition\cite{Stevens65,Ramos71,Frobrich06,Vanherck18} on the commutator Green's functions, yielding the equation 
\begin{equation} \label{eq:AngularCondition}
	\nBmag^{\xmcs}
	= 
		\delta \sinfn{2 \theta},
\end{equation}
which fixes the angle of magnetization $\theta$.
We introduced the normalized magnetic field $\nBmag	\coloneqq \infrac{\landeg \muB \Bmag}{\Mag \texJ}$ and its components $\nBmag^{\xmcs} =\nBmag \sinfn{\Bangle - \theta}$ and $\nBmag^{\zmcs}=	\nBmag \cosfn{\Bangle - \theta}$ in the magnetization reference frame.
Here, equal sublattice magnetization $\Mag\coloneqq \Mag_{\subGen} \equiv \expval{ \OPS{\zmcs}{\sinda} }_{\sinda \in \subGen}$ is assumed, which is reasonable by symmetry and the fact that we are in search for a homogeneous magnetization.
After having removed the singularity, the remaining set of four equations can be solved and used in the spectral theorem.\cite{Tyablikov59,Zubarev60,Tyablikov67}
An additional differential equation needs to be solved,\cite{Callen63} because we are treating the general $S\geqslant\inOver{2}$ case.
All of the above leads to the magnetization
\begin{equation} \label{eq:HigherS}
	\Mag 
	= 
		\frac{ 
			\nbrack{S-\Phi} \nbrack{1+\Phi}^{2S+1} + \nbrack{S+\Phi+1} \Phi^{2S+1} 
		}{ 
			\nbrack{1+\Phi}^{2S+1} - \Phi^{2S+1}
		},
\end{equation}
where in the infinite plane limit $N\to \infty$
\begin{equation} \label{eq:Phi}
	\Phi\opbrack{T} 
	= 
		\Over{\recprimcellvol} \dint{\fbrillzone}{}{ 
			\phi\opbrack{\vk} 
		}{\vk},
\end{equation}
with $\fbrillzone$ the first Brillouin zone (\Fig{fig:honlattice}) and $\recprimcellvol$ the reciprocal primitive cell volume.
The integrand is
\begin{equation} \label{eq:phi}
	\phi 
	=
		\frac{A_{+}}{ \omega_{+} } \cothfn[\Big]{ \frac{\beta}{2} \omega_{+} } 
		-\frac{A_{-}}{ \omega_{-} }\cothfn[\Big]{ \frac{\beta}{2} \omega_{-} } 
		-\Over{2},
\end{equation}
where
\begin{align}
	&{}A_\pm 
	=	
		\frac{ \Mag \exJF\opbrack{0} }{4 \sqrt{\square}} 
		\clbrack[\big]{
			\nbrack{\hcgr + \hcgi} \nbrack{\hcfr \hcgi - \hcfi \hcgr }     
		}\\
		&{}\quad\crbrack[\big]{
			+ \nbrack{\hca - \hcfr -\hcfi} \nbrack{ \pm \sqrt{\square} - \hca\nbrack{\hcfr + \hcfi} - \hcge \nbrack{\hcgr + \hcgi}} 
		},\nonumber\\
	&\square 
	= 
		\nbrack{\hca \hcfr + \hcge \hcgr}^{2}
		+ \nbrack{\hca \hcfi + \hcge \hcgi}^{2}
		- \nbrack{ \hcfr \hcgi - \hcfi \hcgr }^{2}, \nonumber
\end{align}
$\hca	=	
		\nBmag^{\zmcs}	
		+ 1 + \delta \cos{2\theta} 
		- \hcfe$,
$\hcggen=	
			\nanexJF_{\alpha}\opbrack{\vk} \sinsq{\theta}$
and
$\hcfgen=	
		\nexJF_{\alpha}\opbrack{\vk} - \nanexJF_{\alpha}\opbrack{\vk} \cossq{\theta}$
with
$\alpha \in \Set{\Even,\Re,\Im}$.
The quasi-particle excitation energies $\omega_{\pm}$ are defined by
\begin{equation}
	\omega_{\pm}
	\coloneqq 
		\Mag \texJ \sqrt{ 
			\hca^2 + \hcfr^2 + \hcfi^2 
			-\nbrack{\hcge^2 + \hcgr^2 + \hcgi^2} \pm 2 \sqrt{\square} 
		}.
\end{equation}
Materials with a hexagonal lattice can be described by the same solution, considering only interaction with the same sublattice, i.e. $\exJ_1=\exJ_3=0$ (see SI, Sec.~\ref{SI::equalAnis}).

Parameters $\exJ_n$ and $\anis_n$, that the model depends on, were calculated using first-principle calculations in the framework of DFT as implemented in the Vienna \textit{ab-initio} simulation package (VASP).\cite{vasp1,vasp2}
The electron exchange and correlation potentials were described in local density approximation (LDA).\cite{lda-ca}
The Hubbard $U$ term was included as \SI{4}{\electronvolt} for \ch{Cr} and \ch{Mn} to account for the strong on-site Coulomb interaction.\cite{Dudarev} 
The energy cut-off for plane-wave expansion and the energy convergence criteria were set to \SI{500}{\electronvolt} and \SI{e-5}{\electronvolt}, respectively. 
The spin-orbit coupling was included in the non-collinear magnetic calculation. 
In order to extract the magnetic exchange parameters, $\exJ_n$ and $\anis_n$, the four-state energy mapping on Heisenberg spin model was employed.\cite{fourstate} 
The results for three considered monolayer materials (\ch{CrI3}, \ch{CrBr3} and \ch{MnSe2}) are given in Table~\ref{tbl:DFTCurie}.
\begin{table}
	\caption{\label{tbl:DFTCurie} Parametric values obtained using DFT for selected ferromagnetic monolayer materials.}
	\begin{tabularx}{8.5cm}{
			r%
			Y@{.}Z%
			Y@{.}Z%
			Y@{.}Z%
			Y@{.}Z%
			Y@{.}Z%
			Y@{.}Z%
		}
		\toprule
			&\multicolumn{6}{c}{exchange strength $\nbrack{\si{\milli\electronvolt}}$} 	&\multicolumn{6}{c}{anisotropy}
		\\
		\cmidrule(lr){2-7} \cmidrule(l){8-13}
			&\multicolumn{2}{c}{$\exJ_1$}	&\multicolumn{2}{c}{$\exJ_2$}	&\multicolumn{2}{c}{$\exJ_3$}	&\multicolumn{2}{c}{$\anis_1$}	&\multicolumn{2}{c}{$\anis_2$}	&\multicolumn{2}{c}{$\anis_3$}
		\\
		\midrule
			\ch{CrI3}	&3&06	&0&92	&-0&01	&0&08	&-0&05	&-0&92
		\\
			\ch{CrBr3}	&2&72	&0&41	&-0&10	&0&01	&-0&02	&0&05
		\\
			\ch{MnSe2}	&\multicolumn{2}{c}{}	&5&34	&\multicolumn{2}{c}{}	&\multicolumn{2}{c}{}	&0&01	&\multicolumn{2}{c}{}
		\\
		\bottomrule
	\end{tabularx}
\end{table}

Ferromagnetic transition (Curie) temperature $T_{\Curie}$ can be calculated directly from \Eq{eq:HigherS} and \Eq{eq:Phi} in a vanishing external field $\Bmag=0$, and taking the limit $\Mag\to 0$ (details in SI, Sec.~\ref{SI::Curie}).
In agreement with the Mermin-Wagner theorem,\cite{Mermin66} $T_{\Curie}$ vanishes for materials with easy-plane weighted anisotropy $\tnanexJ \leqslant 0$: the in-plane $\mrm{SO}\opbrack{2}$ symmetry is not broken, leading to detrimental quantum fluctuations for any in-plane spontaneous magnetization at finite temperatures.
On the other hand, easy-axis materials ($\tnanexJ>0$) exhibit a finite Curie temperature of
\begin{align} 
	\inv{ T_{\Curie}}
	&=
		\frac{3 \kB}{S\nbrack{S+1} \exJF\opbrack{0}}\Over{\recprimcellvol} 
		\dint{\fbrillzone}{}{ 
			\phi_{\Curie}
		}{\vk} \label{eq:CurieTempGen}
	,\quad\text{with} \\
	\phi_{\Curie}
	&=
		\frac{
			\hca
			+ \nexJF_{\Re}\opbrack{\vk} - \nanexJF_{\Re}\opbrack{\vk}
		}{
			\hca^2
			- \nbrack{\nexJF_{\Re}\opbrack{\vk} - \nanexJF_{\Re}\opbrack{\vk}}^{2}
			- \nbrack{\nexJF_{\Im}\opbrack{\vk} - \nanexJF_{\Im}\opbrack{\vk}}^{2}
		},
\end{align}
where $\hca = 1+\delta -\nexJF_{\Even}\opbrack{\vk}+\nanexJF_{\Even}\opbrack{\vk}$.
In this case, the spontaneous magnetization is out-of-plane, where the continuous rotational symmetry is broken by the anisotropies $\anis_i$.
The Curie temperatures obtained as such for \ch{CrI3}, \ch{CrBr3} and \ch{MnSe2} are in general agreement with experiment and yield a better description than the commonly used Ising model results (see Table~\ref{tbl:CurieGamma}).
Specifically notice the high predicted $T_{\Curie}$ for \ch{MnSe2}, which was recently experimentally observed to exhibit finite spontaneous magnetization at room-temperature by O'Hara \andothers\cite{OHara18}
The remaining difference is potentially due to additional interactions with the substrate or finite-size effects, which are inevitable in experiments and have been shown to result in potentially higher Curie temperatures\cite{Imry69}.
\begin{table}
	\caption{\label{tbl:CurieGamma}
	Comparison of our theoretical results to the existing experimental data for the Curie temperature $T_{\Curie}$ and excitation energies at the $\Gamma$-point $E\opbrack{\Gamma}$ (assuming $\Mag=\Mag_{\Sat}$), for selected ferromagnetic 2D materials.
	The Ising $T_{\Curie}$ is obtained using Monte Carlo simulations (see SI, Sec.~\ref{SI::Curie:MC}).}
	\newcommand*{\thiswork}{theory}
	\newcommand*{\experiment}{experiment}
	\begin{tabularx}{7.75cm}{
			r%
			c%
			c%
			c%
			c%
			c%
		}
		\toprule
			&\multicolumn{3}{c}{$T_{\Curie}$ $\nbrack{\si{\kelvin}}$} 	&\multicolumn{2}{c}{$E\opbrack{\Gamma}$ $\nbrack{\si{\milli\electronvolt}}$}
		\\
		\cmidrule(lr){2-4} \cmidrule(l){5-6}
			&\thiswork &Ising	&\experiment	&\thiswork	&\experiment
		\\
		\midrule
			\ch{CrI3}	&108		&241	&45\cite{Huang17}	&1.46	&2.4\cite{Chen18}
		\\
						&		&		&					&26.6		&19\cite{Chen18}
		\\
			\ch{CrBr3}	&37		&157	&34\cite{Zhang19}, 27\cite{Kim19}	&0.052	&0.1--0.2\cite{Yelon71,Samuelsen71}
		\\
						&		&	&	&23	&15.5\cite{Yelon71}
		\\
		    \ch{MnSe2}	&264	&510	&$\geqslant$RT\cite{OHara18}	&0.96	&-
		\\
		\bottomrule
	\end{tabularx}
\end{table}

\begin{figure}
	\centering
	\includegraphics{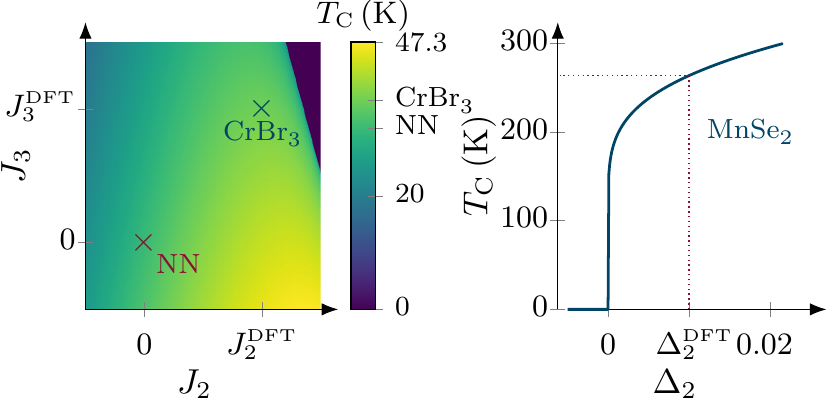}
	\caption{%
		(left) $T_{\Curie}$ as a function of second and third neighbour interaction strengths $\exJ_2$ and $\exJ_3$ for \ch{CrBr3}.
		The values change significantly when accounting for the nearest neighbours only (marked $\mrm{NN}$), compared to further neighbour interactions as calculated from DFT (marked \ch{CrBr3})\@.
		For large $\exJ_2$ and $\exJ_3$, the total weighted anisotropy becomes negative, resulting in a vanishing Curie temperature.
		Similar considerations for \ch{CrI3} show a large dependence on the second neighbour interaction strength.
		(right) $T_{\Curie}$ depends strongly on the anisotropy in \ch{MnSe2} and vanishes for negative anisotropy.%
	}%
	\label{fig:CurieNeighbours}
\end{figure}

In honeycomb materials, second and third neighbour interactions can have a significant influence on the Curie temperature (cf.~\Fig{fig:CurieNeighbours}).
Accounting for nearest-neighbour interactions only in \ch{CrBr3} would result in a Curie temperature of just $\SI{32}{\kelvin}$, compared to $\SI{37}{\kelvin}$ when including further neighbours.
The Curie temperature also strongly depends on the anisotropy, as illustrated for \ch{MnSe2} in \Fig{fig:CurieNeighbours}.
A different combination of further neighbour interactions and anisotropy can even lead to a vanishing Curie temperature if in-plane anisotropy or anti-ferromagnetic interactions dominate.

To facilitate further studies, we supplement this paper with a program\cite{Vanherck20} to calculate Curie temperatures based on \Eq{eq:CurieTempGen}.
The program allows to computationally screen potential 2D ferromagnets solely based on DFT results for their spin, exchange interaction and anisotropy.
Not being limited to materials with relaxed lattice constants, it can also be used to engineer the influence of strain on the Curie temperature.

\begin{figure}
	\centering
	\includegraphics{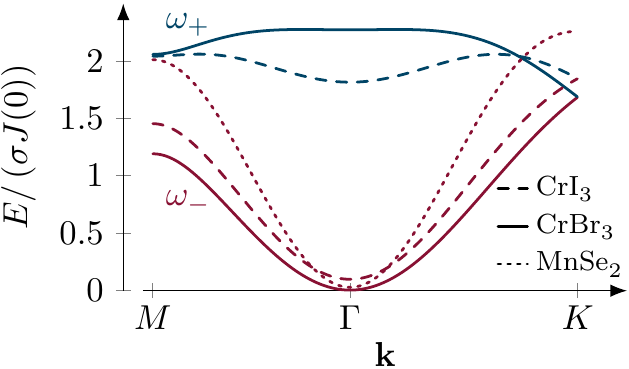}
	\caption{
		Quasi-particle dispersions in spontaneously magnetized state along high-symmetry lines.
		The energy $E=\hbar\omega$ gets renormalized by the relative magnetization $\sigma = \infrac{M\opbrack{T}}{M_{\Sat}}$ at a given temperature $T < T_{\Curie}$.
		The spectra are parabolic with a finite gap at the $\Gamma$-point (listed in Table~\ref{tbl:CurieGamma}).
		Dirac cones typical for the honeycomb materials are present at the $K$-points for \ch{CrI3} and \ch{CrBr3}.
		The single-atomic basis of hexagonal \ch{MnSe2} results only in a single dispersion band.%
	}%
	\label{fig:spectrum}
\end{figure}

Higher Curie temperatures are governed by exciting less quasi-particles, the most basic of which are known as Bloch magnons\cite{Bloch30}, that lower the magnetization at a given temperature.
The energy spectrum $\omega_{\pm}$ for these magnon-like excitations---renormalized by temperature, via spontaneous magnetization $\Mag\opbrack{T}$---is for all materials parabolic with a finite bandgap of $2\Mag \delta \texJ$ in their spontaneously magnetized state (see Fig.~\ref{fig:spectrum}).
It is this finite bandgap, which is in qualitative agreement with available experimental data (Table~\ref{tbl:CurieGamma}), that governs the finite Curie temperature.
The dispersions also exhibit Dirac cones at the $K$-points, which is typical for honeycomb lattice types.\cite{Pershoguba18}
These Dirac cones get gapped when the lattice inversion symmetry is broken, allowing for topological edge states.\cite{Owerre16, Chen18}

\begin{figure}
	\centering
	\includegraphics{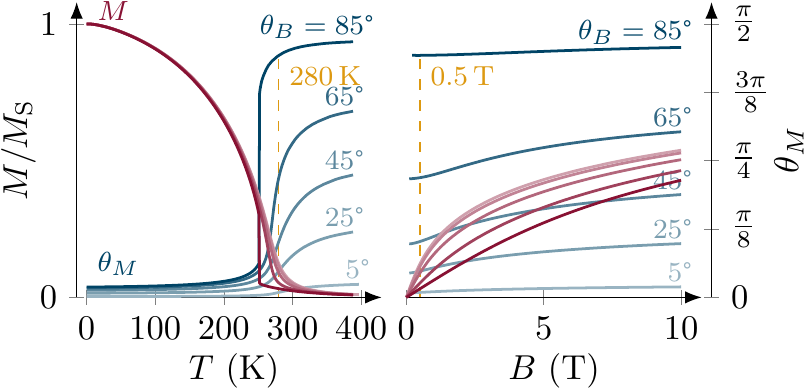}
	\caption{
		Temperature- and applied field-dependence of the relative magnetization $\infrac{M}{M_{\Sat}}$ (left axis) and the magnetization angle $\theta_M$ (right axis) for \ch{MnSe2}.
		The applied field is varied from $\Bangle=\ang{5}$ to $\ang{85}$ relative to the out-of-plane direction.
		The dashed line corresponds to equal temperature and applied field in both panels.
		(left panel) The temperature-dependence is shown for a field of $\Bmag=\SI{0.5}{\tesla}$.
		At low temperatures, the materials anisotropy dominates and the large magnetization is almost out-of-plane.
		Increasing the temperature reduces the magnetization and makes it mostly align with the applied field.
		(right panel) The applied magnetic field is varied at a constant temperature of $\SI{280}{\kelvin}$, slightly above the Curie temperature.
		The materials susceptibility is higher for fields that are applied at an angle close to collinear with the anisotropy.
		Even at substantial fields, the anisotropy prevents the magnetization from fully aligning.%
	}%
	\label{fig:MTAngles}
\end{figure}

Our theory allows for the calculation of both the magnetization angle and magnitude for magnetic fields applied at arbitrary angles $\Bangle$ (\Fig{fig:MTAngles}), which is an important asset in determining magnetic properties of two-dimensional materials.
At weak applied fields, the magnetization remains mostly out of plane at temperatures below the Curie temperature, with only small deviations towards the applied field direction.
Upon raising the temperature or field, the magnetization turns towards the applied field direction, only approaching collinearity in the high field or large temperature limit.
The magnetization behavior is thus dominated by anisotropy at low temperatures and by the external field at high temperatures.
This can be understood from the fact that thermal fluctuations in $\OPvS$ will appear quadratic in $\OPHex$ but only linear in $\OPHB$.
In summary, we presented the theoretical toolbox to characterize the appearance of long-range magnetization in two-dimensional (hexagonal/honeycomb) ferromagnets, under quantum scrutiny.
Using microscopic parameters obtained {\it ab initio}, our quantum Heisenberg model captures behavior of magnetization as a function of temperature and for an arbitrary direction of applied magnetic field, and enables extraction of characteristic excitation (magnon) frequencies in the considered material.
The shown successful demonstration on several materials of present interest validates the methodology, and recommends it for characterizing emergent 2D magnetic materials towards envisaged applications.
Further impact of the method will be reached with further refinements to include e.g.\ dipolar and Dzyaloshinskii-Moriya interactions, which are indispensible for studying skyrmions and other topological excitations.

See supplementary information for calculation and result details.
Data available on request from the authors.
This work was supported by Research Foundation-Flanders (FWO) and the special research funds of the University of Antwerp (BOF-UA).

%
%
%
%
%

%
%
%
%
%
%
%
%
%
%
%
%

%
%
%
%
%
%
%
%
%
%
%
%

%
%
%
%

%
%
\bibliography{paper_apl}
\clearpage

\onecolumngrid
\setcounter{equation}{0}
\setcounter{figure}{0}
\setcounter{table}{0}
\setcounter{page}{1}
\renewcommand{\theequation}{S\arabic{equation}}
\renewcommand{\theHequation}{S\arabic{equation}}
\renewcommand{\thefigure}{S\arabic{figure}}
\renewcommand{\theHfigure}{S\arabic{figure}}
\renewcommand*{\thesection}{S\Roman{section}}
\renewcommand*{\theHsection}{S\Roman{section}}
\renewcommand{\thepage}{S\arabic{page}}
\renewcommand{\thetable}{S\Roman{table}}
\renewcommand{\theHtable}{S\Roman{table}}
\renewcommand{\bibnumfmt}[1]{[S#1]}
\renewcommand{\citenumfont}[1]{S#1}
\makeatletter
{
\begingroup
\frontmatter@title@above
\frontmatter@title@format
Supplementary Information: \ourtitle
\par
\frontmatter@title@below
\endgroup
}
\makeatother

\hfill\begin{minipage}{\dimexpr\textwidth-0.5in}
{\sffamily Joren Vanherck,$^{1,2}$ Cihan Bacaksiz,$^{1,3}$ Bart Sor\'ee,$^{1,2,4}$ Milorad V.\ Milo\v{s}evi\'c,$^{1,3}$ and Wim Magnus$^{1,2}$}\\[.1cm]
  {\itshape ${}^1$\UAaffiliation\\
  ${}^2$\IMECaffiliation\\
  ${}^3$\UAnano\\
  ${}^4$\KULaffiliation\\}
(Dated: \today)\\[0.1cm]
  \end{minipage}

\section{Solution method}
\label{SI::sol}

Here, we present the step-by-step solution for the honeycomb quantum Heisenberg Hamiltonian with anisotropic exchange interactions up to third nearest neighbours with an applied field that can be in arbitrary directions, as detailed in the main paper.
This solution method mostly follows the general framework as set out in Ref.~\citeSI{Vanherck18SI}, to which we refer for more details.

\begin{figure}[b]
	\centering
	\includegraphics{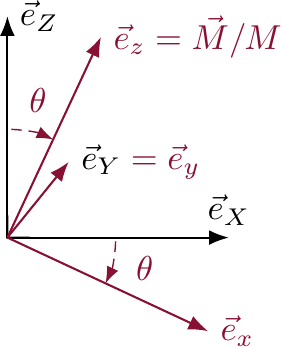}
	\caption{
		\label{fig:coordrot}
		The magnetization reference frame $\Set{\unitvec_\xmcs,\unitvec_\ymcs,\unitvec_\zmcs}$ (red) is obtained by rotating the crystallographic frame of reference $\Set{\unitvec_\xcrys,\unitvec_\ycrys,\unitvec_\zcrys}$ (black) through an angle $\theta$ around the $\unitvec_\ycrys$ axis.%
	}%
\end{figure}

The Hamiltonian $\OPH$ presented in the main text is written with components expressed in terms of the crystallographic coordinate system $\Set{\unitvec_\xcrys,\unitvec_\ycrys,\unitvec_\zcrys}$, defined with $\unitvec_\zcrys$ pointing out of plane, and $\unitvec_\xcrys$ and $\unitvec_\ycrys$ as in \Fig{fig:honlattice}.
We can define a magnetization coordinate system $\Set{\unitvec_\xmcs,\unitvec_\ymcs,\unitvec_\zmcs}$ by requiring $\unitvec_\zmcs\coloneqq \infrac{\vMag}{\Mag}$ to be in the (yet to be determined) direction of magnetization and $\unitvec_\ymcs \coloneqq \unitvec_\ycrys$ (\Fig{fig:coordrot}).
While the Zeeman term $\OPHB$ of the Hamiltonian is invariant under coordinate rotation, with $\infrac{\vBmag}{\Bmag}\equiv \sinfn{\Bangle - \theta}\unitvec_{\xmcs} + \cosfn{\theta - \Bangle} \unitvec_{\zmcs}$ in the magnetization coordinate system, the anisotropic exchange term becomes
\begin{align}
\begin{split}
	\OPHex
	=
		-\frac{\texJ}{2} \smash{\sum_{\sindc} \sum_{\sindd\neq\sindc}}
		& \sqlbrack[\big]{
			\nexJ^{\pmcs\pmcs}_{\sindd\sindc}
			\nbrack{
				\OPS{\pmcs}{\sindc}\OPS{\pmcs}{\sindd}
				+ \OPS{\mmcs}{\sindc}\OPS{\mmcs}{\sindd}
			}
			+ \nexJ^{\zmcs\zmcs}_{\sindd\sindc} \OPS{\zmcs}{\sindc}\OPS{\zmcs}{\sindd}
			}\\& \qquad\sqrbrack[\big]{
			+ \nexJ^{\pmcs\mmcs}_{\sindd\sindc} \OPS{\pmcs}{\sindc}\OPS{\mmcs}{\sindd}
			+ \nexJ^{\pmcs\zmcs}_{\sindd\sindc}
			\nbrack{
				\OPS{\pmcs}{\sindc}\OPS{\zmcs}{\sindd}
				+ \OPS{\mmcs}{\sindc}\OPS{\zmcs}{\sindd}
			}
		},
\end{split}
\end{align}
with raising and lowering operators $\OPS{\pmmcs}{\sindc} \coloneqq \OPS{\xmcs}{\sindc} \pm \OPS{\ymcs}{\sindc}$ and normalized exchange tensor components
\begin{gather}
\begin{aligned}
	\nexJ^{\pmcs\pmcs}_{\sindd\sindc}
	&\coloneqq
		\Over{2} \nanexJ_{\sindd\sindc} \sinsq{\theta}
	\\
	\nexJ^{\zmcs\zmcs}_{\sindd\sindc}
	&\coloneqq
		 \nexJ_{\sindd\sindc} + \nanexJ_{\sindd\sindc}\cos{2\theta}
	\\
	\nexJ^{\pmcs\mmcs}_{\sindd\sindc}
	&\coloneqq
		 \nexJ_{\sindd\sindc} - \nanexJ_{\sindd\sindc}\cossq{\theta}
	\\
	\nexJ^{\pmcs\zmcs}_{\sindd\sindc}
	&\coloneqq
		-\nanexJ_{\sindd\sindc}\sin{2\theta},
\end{aligned}
\end{gather}
where $\theta$ ($0\leqslant \theta \leqslant \inpitwo$) is the angle between $\unitvec_\zmcs$ and $\unitvec_\zcrys$.

We are interested in calculating the homogeneous, stationary sublattice magnetization $\Mag_\subGen$ in the $\zmcs$-direction, which can be quantified by the expectation value of the $\zmcs$-component of the spins in sublattice $\subGen\in \Set{\subA,\subB}$:
\begin{equation}
	\Mag_{\subGen}
	\coloneqq
		\expval{ \OPS{\zmcs}{\sinda} }
	,\quad \sinda\in \subGen
\end{equation}
with the canonical ensemble average
\begin{equation} \label{eq:Green:CanAvg}
\corrf{\dots} \coloneqq \inv{Z}\Tr\nbrack{\expof{-\beta\OPH}\dots},\qquad Z\coloneqq \Tr\nbrack{\expof{-\beta\OPH}}
\end{equation}
at temperature $T$ ($\beta\coloneqq\inOver{\kBT}$).
Since we are looking for a homogeneous quantity, the expectation value itself is independent of the lattice site $\sinda$ (because of the equivalence of all lattice sites of the sublattice).
Due to the material's symmetry, it is furthermore reasonable to assume the sublattice magnetizations to be equal $\Mag\coloneqq \Mag_{\subA} \equiv \Mag_{\subB}$.
Next to the magnetization magnitude $\Mag$, we also need to determine its direction $\theta$.

We define the first and second-order Green's functions\citeSI{Zubarev60SI} in the frequency domain
\begin{subequations}
	\begin{align}
	\Green^{\subGen,\alpha}_{\sinda,\sindb}\opbrack{\omega, \lambda}
	&\coloneqq
		\greenf{
			\OPS{\alpha}{\sinda}
		}{
			\OPO_\sindb\opbrack{\lambda}
		}\opbrack{\omega}
		,\quad \sinda\in \subGen
	\\
	\Green^{\subGen\subGenp,\alpha\beta}_{\sinda\sindd,\sindb}\opbrack{\omega,\lambda}
	&\coloneqq
		\greenf{
			\OPS{\alpha}{\sinda}\OPS{\beta}{\sindd}
		}{
			\OPO_\sindb\opbrack{\lambda}
		}\opbrack{\omega}
		,\quad \sinda \in \subGen,\sindd \in \subGenp \label{eq:GreenSO}
	\end{align}
\end{subequations}
with $\alpha,\beta \in \Set{\pmcs,\mmcs,\zmcs}$, as
\begin{equation}
	\greenf{
		\OPA
	}{
		\OPB
	}\opbrack{\omega}
	\coloneqq
		\Over{2\pi} \dint{-\infty}{\infty}{
			\greenf{
				\OPA\opbrack{t}
			}{
				\OPB\opbrack{0}
			}
			\expof{\ii \omega t}
		}{t}
\end{equation}
with the Green's function in time domain being defined as
\begin{equation}
	\greenf{
		\OPA\opbrack{t}
	}{
		\OPB\opbrack{0}
	}
	\coloneqq
		-\ii \theta\opbrack{t} \corrf{\commutator{\OPA\opbrack{t}}{\OPB\opbrack{0}}},
\end{equation}
where the quantum mechanical operators are in the Heisenberg representation, $\commutator{\OPA}{\OPB}$ denotes a commutator and $\theta\opbrack{t}$ is the Heaviside step function.
The operator
\begin{equation}
	\OPO_\sindb\opbrack{\lambda}
	\coloneqq
		\expof{ \lambda \OPS{\zmcs}{\sindb} } \OPS{\mmcs}{\sindb}
\end{equation}
contains a real parameter $\lambda$, necessary\citeSI{Callen63SI} to treat cases with $S>\inOver{2}$.
The symbols $\subGen$ and $\subGenp$ denote either one of the sublattices $\subA$ or $\subB$.

The equation of motion obeyed by the first order Green's functions is
\begin{equation}
	\omega \Green^{\subGen,\alpha}_{\sinda,\sindb}
	=
		\Over{2\pi} \expval{
			\commutator{\OPS{\alpha}{\sinda}}{\OPO_\sindb}
		} \delta_{\sinda\sindb}
		+ \greenf{
			\commutator{\OPS{\alpha}{\sinda}}{\OPH}
		}{
			\OPO_\sindb
		}
		,\quad \sinda\in \subGen,
\end{equation}
where we dropped the explicit frequency- and $\lambda$-dependence.
The commutator in the last term introduces second order Green's functions of the form \Eq{eq:GreenSO}
which can be decoupled using the RPA\citeSI{Englert60SI} (Tyablikov\citeSI{Tyablikov59SI}) decoupling approximation
\begin{equation}
	\Green^{\subGen\subGenp,\alpha\beta}_{\sinda\sindd,\sindb}
	\rightarrow
		\expval{ \OPS{\alpha}{\sinda} } \Green^{\subGenp,\beta}_{\sindd,\sindb}
		+ \expval{ \OPS{\beta}{\sindd} } \Green^{\subGen,\alpha}_{\sinda,\sindb}
	,\quad \sinda \in \subGen,\sindd \in \subGenp.
\end{equation}
Notice that the commutator with the Hamiltonian couples the \subA{} and \subB{} sublattices through the first and third nearest neighbor exchange interactions, resulting in a set of coupled equations that is double the size of that for a material with a single atom basis.
By definition of the magnetization coordinate system, we know
\begin{equation}
\begin{aligned}
	\expval{ \OPS{\pmcs}{\sinda} }
	=
		\expval{ \OPS{\mmcs}{\sinda} }
	&\equiv
		0
	\\
	\expval{ \OPS{\zmcs}{\sinda} }
	&\equiv
		\Mag_{\subGen},\quad \sinda \in \subGen
\end{aligned}
\end{equation}
simplifying the resulting set of equations significantly.

Next, we perform a spatial Fourier transform to exploit the lattice symmetry.
Assuming the lattice has $N$ primitive cells, all with a two atom basis, we define
\begin{subequations}
\begin{align}
	\Green^{\subGen,\alpha}_{\sinda,\sindb}\opbrack{\omega}
	&\coloneqq
		\Over{N} \sum_{\vk\in\fbrillzone} \expof{-\ii\vk\bcdot\nbrack{\sindbpos-\sindapos}} \Green^{\subGen,\alpha}\opbrack{\omega,\vk}
	\\
	\delta_{\sinda\sindb}
	&\equiv
		\Over{N} \sum_{\vk\in\fbrillzone} \expof{-\ii\vk\bcdot\nbrack{\sindbpos-\sindapos}}
	\\
	\nexJF^{\alpha\beta}_\subGen\opbrack{\vk}
	&\coloneqq
		\sum_{\lvec{\sindd\sinda}} \nexJ^{\alpha\beta}_{\sindd\sinda} \expof{\ii\vk\bcdot\lvec{\sindd\sinda}}
	=
		\sum_{\sindd} \nexJ^{\alpha\beta}_{\sindd\sinda} \expof{\ii\vk\bcdot\nbrack{\sinddpos-\sindapos}},
\end{align}
\end{subequations}
where $\lvec{\sindd\sinda} \coloneqq \sinddpos-\sindapos$ and $\sinda$ is located on sublattice $\subGen$.
This means that $\lvec{\sindd\sinda}$ is a vector that points from a lattice point on sublattice $\subGen$ towards some other lattice point.
After the spatial Fourier transform, the equations of motion (without explicitly writing the $\vk$-dependence) can be written as
\begin{equation}
	\nbrack{\omega \one - \vGamma}\bcdot\vGreen
	=
		\Over{2\pi} \vPsi\opbrack{\lambda},
\end{equation}
with the Green's function vector
\begin{equation}
	\vGreen
	\coloneqq
		\begin{bmatrix}
			\vGreen^{\subA} \\
			\vGreen^{\subB}
		\end{bmatrix}
	\text{, where } \quad
	\vGreen^{\subGen}
	\coloneqq
		\begin{bmatrix}
			\Green^{\subGen,\pmcs} \\
			\Green^{\subGen,\mmcs} \\
			\Green^{\subGen,\zmcs}
		\end{bmatrix}.
\end{equation}
Terms containing $\vGreen^{\subA}$ have two origins: (i) the Green's function equation of motion of the \subA{} sublattice; and (ii) the first and third nearest neighbor interaction in the equation of motion of the \subB{} sublattice (and vice versa).
We defined an inhomogeneity vector
\begin{equation}
	\vPsi\opbrack{\lambda}
	\coloneqq
		\begin{bmatrix}
			\vPsi^{\subA}\opbrack{\lambda} \\
			\vPsi^{\subB}\opbrack{\lambda}
		\end{bmatrix},
\end{equation}
where
\begin{equation}
	\vPsi^{\subGen}\opbrack{\lambda}
	\coloneqq
		\begin{bmatrix}
			\expval{
				\commutator{
					\OPS{\pmcs}{\sinda}
				}{
					\expof{ \lambda \OPS{\zmcs}{\sinda} }\OPS{\mmcs}{\sinda}
				}
			}
		\\
			\expval{
				\commutator{
					\OPS{\mmcs}{\sinda}
				}{
					\expof{ \lambda \OPS{\zmcs}{\sinda} }\OPS{\mmcs}{\sinda}
				}
			}
		\\
			\expval{
				\commutator{
					\OPS{\zmcs}{\sinda}
				}{
					\expof{ \lambda \OPS{\zmcs}{\sinda} }\OPS{\mmcs}{\sinda}
				}
			}
		\end{bmatrix}
	=
		\begin{bmatrix}
			\vPsi^{\subGen,\pmcs}\opbrack{\lambda} \\
			0 \\
			0
		\end{bmatrix}
	,\quad \sinda\in \subGen.
\end{equation}
The coefficient matrix $\vGamma$ has two eigenvalues---one for each sublattice $\subGen$---that are vanishing for all $\vk$.
If we denote the corresponding left eigenvectors by $\vL_{0,\subGen}$, the regularity condition for commutator Green's functions\citeSI{Stevens65SI,Ramos71SI,Frobrich06SI,Vanherck18SI} yields
\begin{equation}
	\vL_{0,\subGen} \bcdot \vPsi\opbrack{\lambda}
	=
		0
\end{equation}
for $\subGen\in\Set{\subA, \subB}$.
Both result in the same angular condition \eqref{eq:AngularCondition}, fixing the angle of magnetization $\theta$ for a given magnetization magnitude.
Using these regularity conditions in $\vGamma$, implies
\begin{equation}
	\Green^{\subGen,\zmcs}
	=
		0.
\end{equation}
This effectively reduces $\vGamma$ to a $4\times 4$ matrix, by eliminating the third and sixth rows and columns.
From here on, we will always refer to this reduced set of equations as the equations of motion.

The four eigenvalues of the reduced $\vGamma$ are
\begin{equation}
	\omega_{\opm \pm}
	=
		\opm \Mag \exJF\opbrack{0} \sqrt{
			\hca^2 + \hcfr^2 + \hcfi^2
			- \nbrack{\hcge^2 + \hcgr^2 + \hcgi^2} \pm 2 \sqrt{\square}
		},
\end{equation}
where the circle around the first $\pm$ sign is added to distinguish the labels.
In the above equation, we defined
\begin{eqnarray}
	\square
	\coloneqq
		\hca^2 \nbrack{ \hcfr^2 + \hcfi^2 }
		+ 2 \hca \hcge \nbrack{\hcfr \hcgr + \hcfi \hcgi}
		\qquad\\
		{}+ \hcge^2 \nbrack{\hcgr^2 + \hcgi^2}
		- \nbrack{ \hcfr \hcgi - \hcfi \hcgr }^{2}, \nonumber
\end{eqnarray}
and
\begin{subequations}
\begin{align}
	\hca
	&\coloneqq
		\nBmag^{\zmcs}
		+ \nexJF^{\zmcs\zmcs}\opbrack{0}
		- \nexJF^{\pmcs\mmcs}_{\Even}\opbrack{\vk}
	=
		\nBmag^{\zmcs}
		+ 1 + \delta \cos{2\theta}
		- \hcfe
	\\
	\hcfe
	&\coloneqq
		\nexJF^{\pmcs\mmcs}_{\Even}\opbrack{\vk}
		=
		\nexJF_{\Even}\opbrack{\vk} - \nanexJF_{\Even}\opbrack{\vk} \cossq{\theta}
	\\
	\hcfr
	&\coloneqq
		\nexJF^{\pmcs\mmcs}_{\Re}\opbrack{\vk}
		=
		\nexJF_{\Re}\opbrack{\vk} - \nanexJF_{\Re}\opbrack{\vk} \cossq{\theta}
	\\
	\hcfi
	&\coloneqq
		\nexJF^{\pmcs\mmcs}_{\Im}\opbrack{\vk}
	=
		\nexJF_{\Im}\opbrack{\vk} - \nanexJF_{\Im}\opbrack{\vk} \cossq{\theta}
	\\
	\hcge
	&\coloneqq
		2 \nexJF^{\pmcs\pmcs}_{\Even}\opbrack{\vk}
		=
		\nanexJF_{\Even}\opbrack{\vk} \sinsq{\theta}
	\\
	\hcgr
	&\coloneqq
		2 \nexJF^{\pmcs\pmcs}_{\Re}\opbrack{\vk}
	=
		\nanexJF_{\Re}\opbrack{\vk} \sinsq{\theta}
	\\
	\hcgi
	&\coloneqq
		2 \nexJF^{\pmcs\pmcs}_{\Im}\opbrack{\vk}
	=
		\nanexJF_{\Im}\opbrack{\vk} \sinsq{\theta}.
\end{align}
\end{subequations}
Each of the eigenvalues $\omega_{\opm \pm}$ correspond with a left eigenvector $\vL_{\opm \pm}$ and a right eigenvector $\vR_{\opm \pm}$, which can be combined in matrices as
\begin{equation}
	\vL
	=
		\begin{bmatrix}
			\vL_{\ominus -}
		\\
			\vL_{\oplus -}
		\\
			\vL_{\ominus +}
		\\
			\vL_{\oplus +}
		\end{bmatrix}
	\text{, and }
	\vR
	=
		\begin{bmatrix}
			\vR_{\ominus -}
		; &
			\vR_{\oplus -}
		; &
			\vR_{\ominus +}
		; &
			\vR_{\oplus +}
		\end{bmatrix}.
\end{equation}
such that they are properly normalized, orthonormal and thus $\vL \bcdot \vR = \vR \bcdot \vL = \one$, while
\begin{equation}
	\vL \bcdot \vGamma \bcdot \vR
	=
		\vOmega
	\coloneqq
		\begin{bmatrix}
			\omega_{\ominus -}	&	0	&	0	&	0 	\\
			0	&	\omega_{\oplus -}	&	0	&	0 	\\
			0	&	0	&	\omega_{\ominus +}	&	0 	\\
			0	&	0	&	0	&	\omega_{\oplus +}
		\end{bmatrix}.
\end{equation}
The $n^{\text{th}}$ element of eigenvector $\vL_{\opm \pm}$ will be written as $L_{\opm \pm}^{\nbrack{n}}$, and similar for the right eigenvectors.

Using the orthonormality of the left and right eigenvectors, we can write the Green's function equations of motion as
\begin{equation}
	\nbrack{\omega\one - \vOmega} \vtGreen
	=
		\Over{2\pi} \vL \vPsi\opbrack{\lambda}.
\end{equation}
with
\begin{equation}
	\vtGreen
	\coloneqq
		\vL \vGreen,
	\text{ and }
	\vPsi\opbrack{\lambda}
	=
		\begin{bmatrix}
				\Psi^{\subA}\opbrack{\lambda}
			\\[2ex]
				0
			\\[2ex]
				\Psi^{\subB}\opbrack{\lambda}
			\\[2ex]
				0
		\end{bmatrix}
	\coloneqq
		\begin{bmatrix}
				\Psi^{\subA,\pmcs}\opbrack{\lambda}
			\\[2ex]
				0
			\\[2ex]
				\Psi^{\subB,\pmcs}\opbrack{\lambda}
			\\[2ex]
				0
		\end{bmatrix}
\end{equation}
We can now solve the Green's function equation of motion.
The transformed inhomogeneity vector is
\begin{equation}
	\vL \vPsi\opbrack{\lambda}
	=
		\vL^{\nbrack{1}} \Psi^{\subA}
		+ \vL^{\nbrack{3}} \Psi^{\subB}.
\end{equation}
Solving this transformed equation of motion for
\begin{equation}
	\vtGreen
	=
		\begin{bmatrix}
			\tGreen_{\ominus \mmcs}
		\\
			\tGreen_{\oplus \mmcs}
		\\
			\tGreen_{\ominus \pmcs}
		\\
			\tGreen_{\oplus \pmcs}
		\end{bmatrix},
\end{equation}
we find
\begin{equation}
	\tGreen_{\opm \pm}
	=
		\Over{2 \pi} \Over{\omega - \omega_{\opm \pm} }
		\nbrack{
			L_{\opm \pm}^{\nbrack{1}} \Psi^{\subA}
			+ L_{\opm \pm}^{\nbrack{3}} \Psi^{\subB}
		}.
\end{equation}
We now define the transformed expectation values
\begin{equation}
	\vtcorr_\lambda
	\coloneqq
		\vL\vcorr_\lambda
	=
		\begin{bmatrix}
			\tcorr_{\lambda}^{\nbrack{1}}
		\\[1ex]
			\tcorr_{\lambda}^{\nbrack{2}}
		\\[1ex]
			\tcorr_{\lambda}^{\nbrack{3}}
		\\[1ex]
			\tcorr_{\lambda}^{\nbrack{4}}
		\end{bmatrix},
\end{equation}
based on the expectation values
\begin{equation}
	\vcorr_\lambda
	\coloneqq
		\begin{bmatrix}
			\corr_{\lambda}^{\subA,\pmcs}
		\\[1ex]
			\corr_{\lambda}^{\subA,\mmcs}
		\\[1ex]
			\corr_{\lambda}^{\subB,\pmcs}
		\\[1ex]
			\corr_{\lambda}^{\subB,\mmcs}
		\end{bmatrix}
	=
		\begin{bmatrix}
			\expval{
				\expof{\lambda\OPS{\zmcs}{}}\OPS{\mmcs}{}\OPS{\pmcs}{}
			}_{\subA} \opbrack{\vk}
		\\[1ex]
			\expval{
				\expof{\lambda\OPS{\zmcs}{}}\OPS{\mmcs}{}\OPS{\mmcs}{}
			}_{\subA} \opbrack{\vk}
		\\[1ex]
			\expval{
				\expof{\lambda\OPS{\zmcs}{}}\OPS{\mmcs}{}\OPS{\pmcs}{}
			}_{\subB} \opbrack{\vk}
		\\[1ex]
			\expval{
				\expof{\lambda\OPS{\zmcs}{}}\OPS{\mmcs}{}\OPS{\mmcs}{}
			}_{\subB} \opbrack{\vk}
		\end{bmatrix}.
\end{equation}
We now use the spectral theorem\citeSI{Tyablikov59SI,Zubarev60SI,Tyablikov67SI} together with the expression for $\vtGreen$ to find
\begin{equation}
	\tcorr_{\opm \pm}
	=
		\nu_{\opm \pm} \nbrack{
			L_{\opm \pm}^{\nbrack{1}} \Psi^{\subA}
			+ L_{\opm \pm}^{\nbrack{3}} \Psi^{\subB}
		}
\end{equation}
with
\begin{equation}
	\nu_{\opm \pm}
	=
		\Over{ \expof{\beta \omega_{\opm \pm} } -1 }
\end{equation}
for the transformed expectation values.
As we did before, it is a reasonable assumption that the inhomogeneity terms for both sublattices are equal,
\begin{equation}
	\Psi
	\coloneqq
		\Psi^{\subA}
	\equiv
		\Psi^{\subB},
\end{equation}
such that the transformed expectation values are given by
\begin{equation}
	\tcorr_{\opm \pm}
	=
		\nu_{\opm \pm} \nbrack{
			 L_{\opm \pm}^{\nbrack{1}}
			 + L_{\opm \pm}^{\nbrack{3}}
		} \Psi.
\end{equation}
Here, we are interested in calculating $\corr_{\lambda}^{\subA,\pmcs}$ (analogous calculations with $\corr_{\lambda}^{\subB,\pmcs}$ gives the same result), which we can do as
\begin{equation}
	\corr_{\lambda}^{\pmcs}
	\coloneqq
		\corr_{\lambda}^{\subA,\pmcs}
	=
		\vR^{\nbrack{1}} \bcdot \vtcorr.
\end{equation}
After elimination of some common factors, this can be written as
\begin{equation}
	\corr_{\lambda}^{\pmcs}
	=
		\phi \Psi,
\end{equation}
where we defined
\begin{equation}
	\phi
	\coloneqq
		\Over{4 \sqrt{\square} }
		\nbrack[\Big]{
			-\frac{\nu_{\ominus -}}{\omega_{\ominus -}} Q_{\ominus -}
			- \frac{\nu_{\oplus -}}{\omega_{\oplus -}} Q_{\oplus -}
			+ \frac{\nu_{\ominus +}}{\omega_{\ominus +}} Q_{\ominus +}
			+ \frac{\nu_{\oplus +}}{\omega_{\oplus +}} Q_{\oplus +}
		}
\end{equation}
with the combination
\begin{equation}
	Q_{\opm \pm}
	\coloneqq
		\Realpart \opbrack[\big]{
			L_{\opm \pm}^{\nbrack{1}}
			+ L_{\opm \pm}^{\nbrack{3}}
		}.
\end{equation}
In this last equation, we dropped the imaginary part of $L_{\opm \pm}^{\nbrack{1}} + L_{\opm \pm}^{\nbrack{3}}$.
This is allowed since we will later integrate $\corr_{\lambda}^{\pmcs}$ over the first Brillouin zone.
The imaginary part is however odd in $k_\xcrys$, such that it would vanish anyway after integration.
Notice that $\omega_{\ominus \pm} = -\omega_{\oplus \pm}$, such that we can use the shorthand
\begin{equation}
	\omega_{\pm}
	\coloneqq
		\omega_{\oplus \pm}
	\equiv
		-\omega_{\ominus \pm}.
\end{equation}
In order to get sensible results, we moreover require $\omega_{\pm}^2 \geqslant 0$.
The equation for $\phi$ can then be written as
\begin{equation}
	\phi
	=
		\Over{4 \sqrt{\square} }
		\nbrack[\Big]{
			\Over{\omega_{+}}
			\nbrack{
				\nu_{\oplus +} Q_{\oplus +}
				- \nu_{\ominus +} Q_{\ominus +}
			}
			- \Over{\omega_{-}}
			\nbrack{
				\nu_{\oplus -} Q_{\oplus -}
				- \nu_{\ominus -} Q_{\ominus -}
			}
		}
\end{equation}
and using the combinations
\begin{subequations}
\begin{align}
	\nu_{\oplus \pm} + \nu_{\ominus \pm}
	&=
		-1
	\\
	\nu_{\oplus \pm} - \nu_{\ominus \pm}
	&=
		\cothfn[\Big]{\frac{\beta}{2}\omega_{\pm}}
\end{align}
\end{subequations}
this results in \Eq{eq:phi}.

We now found $\corr_{\lambda}^{\pmcs}$ in $\vk$-space, allowing us to calculate
\begin{equation}
	\exval_{\sinda\sinda,\lambda}
	\coloneqq
		\expval{
			\expof{ \lambda \OPS{\zmcs}{\sinda} } \OPS{\mmcs}{\sinda} \OPS{\pmcs}{\sinda}
		}
	=
		\Over{N} \sum_{\vk\in\fbrillzone} \corr_{\lambda}^{\pmcs}
	=
		\Phi \Psi\opbrack{\lambda}
\end{equation}
in direct space, where we introduced
\begin{subequations} \label{eq:SIPhi}
\begin{align}
	\Phi\opbrack{T}
	&\coloneqq
		\Over{N}\sum_{\vk\in\fbrillzone} \phi\opbrack{\vk}
	\\&\approx
		\Over{\recprimcellvol} \dint{\fbrillzone}{}{
			\phi\opbrack{\vk}
		}{\vk}.
\end{align}
\end{subequations}
The last approximation, with $\recprimcellvol$ the primitive reciprocal cell volume, becomes exact as $N\to\infty$.
This factor $\Phi\opbrack{T}$ can take all positive values and vanishes when all spins are aligned (saturation magnetization).
From here on, one can follow Callen~\citeSI{Callen63SI} to find the additional relation \Eq{eq:HigherS}.

\section{Curie temperature}
\label{SI::Curie}
\subsection{Analytical}
\label{SI::Curie:Analytical}

We are now in a position to calculation the Curie temperature.
The Curie temperature $T_{\Curie}$ is defined as the temperature at which $\Mag\to 0$ as $T\to T_{\Curie}$ in the absence of an external magnetic field (put $\Bmag=0$).
Notice from \Eq{eq:HigherS} that $\Phi \to \infty$ as $\Mag\to 0$, allowing that formula to be expanded in therms of $\inv{\Phi}\to 0$, giving the relationship
\begin{equation} \label{eq:SICurieInvMag}
	\inv{\Mag}
	\overset{\Mag\to 0}{\approx}
		\frac{3}{S\nbrack{S+1}}\Phi
\end{equation}

Now notice that, as $\Mag\to 0$, $\omega_{\pm}\to 0$, allowing for the expansion $\coth{x} \overset{x\to 0}{\approx}\inOver{x}$ in \Eq{eq:phi}:
\begin{equation} \label{eq:SIphiExpanded}
	\phi
	\overset{\Mag\to 0}{\approx}
		\frac{1}{\beta \Mag \exJF\opbrack{0}}
		\phi_{\Curie}
		-\Over{2},
\end{equation}
with
\begin{equation}
	\phi_{\Curie}
	\coloneqq
		2\frac{
			\infrac{A_{+}}{ \nbrack{\Mag \exJF\opbrack{0}} }
		}{
			\infrac{\omega_{+}^2 }{ \nbrack{\Mag \exJF\opbrack{0}}^{2} }
		}
		-2\frac{
			\infrac{A_{-}}{ \nbrack{\Mag \exJF\opbrack{0}} }
		}{
			\infrac{\omega_{-}^2 }{ \nbrack{\Mag \exJF\opbrack{0}}^{2} }
		}.
\end{equation}
Note that $\phi_{\Curie}$ is independent of $\Mag$, such that all $\Mag$-dependence is explicit in the formula for $\phi$.
This means that as $\Mag\to 0$, we can neglect the last term.
Together with \Eq{eq:SICurieInvMag} and \Eq{eq:SIPhi}, we find, after eliminating the remaining factor $\inOver{\Mag}$, the Curie temperature $T_{\Curie}$ as
\begin{equation} \label{eq:SICurieTempGen}
	\Over{\kB T_{\Curie}}
	=
		\frac{3}{S\nbrack{S+1} \exJF\opbrack{0}}\Over{\recprimcellvol}
		\dint{\fbrillzone}{}{
			\phi_{\Curie}
		}{\vk}.
\end{equation}

In order to sustain a finite magnetization in the absence of a magnetic field, the angular condition \eqref{eq:AngularCondition} requires $\theta=0,\inpitwo$.
The choice $\theta=\inpitwo$ leads to a two-dimensional integral that diverges to infinity in \Eq{eq:SICurieTempGen}, resulting in $T_{\Curie}=0$.
We thus put $\theta=0$ from now on.
The integrand can then be written as
\begin{equation}
	\phi_{\Curie}
	=
		\frac{
			\hca
			+ \nexJF_{\Re}\opbrack{\vk} - \nanexJF_{\Re}\opbrack{\vk}
		}{
			\hca^2
			- \nbrack{\nexJF_{\Re}\opbrack{\vk} - \nanexJF_{\Re}\opbrack{\vk}}^{2}
			- \nbrack{\nexJF_{\Im}\opbrack{\vk} - \nanexJF_{\Im}\opbrack{\vk}}^{2}
		},
\end{equation}
where
\begin{equation}
	\hca
	=
		1+\delta -\nexJF_{\Even}\opbrack{\vk}+\nanexJF_{\Even}\opbrack{\vk}.
\end{equation}
This yields a finite positive Curie temperature if and only if the weighted anisotropy $\delta$ is positive.
The latter corresponds to having non-negative quasi-particle excitation energies $\omega_{\pm} \geqslant 0$.
Physically, it also makes sense that a spontaneous out-of-plane magnetization is only possible with an overall easy-axis anisotropy $\delta$.

\subsection{Ising Monte-Carlo}
\label{SI::Curie:MC}
For additional check-point in validating temperature-dependent magnetization of the considered 2D materials, we also performed Monte-Carlo simulations based on Ising model.
We considered 2048-atom $32 \times 32$  and 1600-atom $40 \times 40$ supercell for \ch{CrX3} (\ch{X}= \ch{I}, \ch{Br}) and \ch{MnSe2}, respectively.
$2 \times 10^3$ spin-flip steps per site were considered to reach the thermal equilibrium.
In Table \ref{mc-ising}, the magnetic exchange parameters and corresponding critical temperatures are listed.
\begin{table}[htbp]
\caption{
	Magnetic exchange parameters and the corresponding critical temperatures based on the Ising Model for monolayer \ch{CrI3}, \ch{CrBr3} and \ch{MnSe2}.
	\ch{Cr} and \ch{Mn} atoms form two-dimensional honeycomb (hc)  and triangular (tri) lattices. $\exJ^{\xcrys\xcrys}$, $\exJ^{\ycrys\ycrys}$, and $\exJ^{\zcrys\zcrys}$ describe the magnetic interaction between neighboring sites in the Heisenberg spin model.
	The three dimensions interaction was reduced to Ising approximation by averaging the parameters as $\exJ^{\text{Is}}= \infrac{\nbrack{\exJ^{\xcrys\xcrys}+\exJ^{\ycrys\ycrys}+\exJ^{\zcrys\zcrys}}}{3}$.
	The respective critical temperatures obtained using MC simulations are listed in $T_{\Curie}^{\text{MC-Is}}$ column. $T_{\Curie}^{\text{Ex-Is}}$ values are obtained using exact solutions of the Ising model for two-dimensional honeycomb and triangular lattices where the formulae are $T_{\Curie}^{ \text{hc}}=\frac{2S^2}{\ln\opbrack{2+\sqrt{3}}}\frac{\exJ^{\text{Is}}}{\kB}$ and $ T_{\Curie}^{ \text{tri}}=\frac{4S^2}{\ln\opbrack{3}}\frac{\exJ^{\text{Is}}}{\kB}$, respectively, where $S=\infrac{3}{2}$.
	Note that the formula for honeycomb lattice is valid only for the 1st NN approximation, therefore we also listed the MC results for 1st NN in parenthesis. }
\label{mc-ising}
\begin{tabular}{lcccccccc}
\toprule
     &Lattice & $\exJ^{\xcrys\xcrys}$ & $\exJ^{\ycrys\ycrys}$ & $\exJ^{\zcrys\zcrys}$  & $\exJ^{\text{Is}}$ & $T_{\Curie}^{\text{MC-Is}}$ & $T_{\Curie}^{\text{Ex-Is}}$   \\
     &        & $\nbrack{\si{\milli\electronvolt}}$    &   $\nbrack{\si{\milli\electronvolt}}$  &    $\nbrack{\si{\milli\electronvolt}}$      &    $\nbrack{\si{\milli\electronvolt}}$ &    $\nbrack{\si{\kelvin}}$        &   $\nbrack{\si{\kelvin}}$   \\
\midrule
\ch{CrI3} & Honeycomb  &       &          &                 &          &        241 (120)      &    117.8    \\
1st NN  &            & 2.81   & 2.81     &  3.30            & 2.97     &               &       \\
2nd NN  &            & 0.97   & 0.97     &  0.87          & 0.94     &               &       \\
3rd NN  &            & -0.03 & -0.03   & -0.00(1)          & -0.02    &               &        \\
\midrule
\ch{CrBr3} & Honeycomb  &       &         &                     &           &      157  (106)       &     107.5    \\
1st NN  &             & 2.70    & 2.70    &   2.74        &  2.71     &              &       \\
2nd NN  &             & 0.42    & 0.42    &   0.40        &  0.41     &              &        \\
3rd NN  &             & -0.09   & -0.09   &   -0.10       & -0.10     &              &       \\
\midrule
\ch{MnSe2} & Triangular  &       &          &                     &         &        510   &  506.1     \\
1st NN   &             & 5.30  &  5.30    &  5.37          & 5.32    &              &           \\
\bottomrule
\end{tabular}
\end{table}

\section{Equal anisotropies}
\label{SI::equalAnis}

In this section, we explicitly show the limiting case where the anisotropies are all equal, $\anis \equiv \anis_i$.
This is useful limit, since it avoids instabilities in the numerical calculations on the one hand, while on the other hand it reduces the parameter space to get some more useful insights.

With this constraint of equal anisotropies, the variables $A_\pm$ in the integrand \eqref{eq:phi} become
\begin{equation}\label{eq:SIApmEqualAnis}
	A_\pm
	=
		\frac{\Mag \exJF\opbrack{0}}{ 4 } \nbrack{\pm 1-\alpha} t_{\pm}
\end{equation}
and the quasi-particle dispersions
\begin{equation}
	\omega_{\pm}
	=
		\Mag \exJF\opbrack{0} \sqrt{t_{\pm}^2-p_{\pm}^2},
\end{equation}
where we introduced
\begin{align}
	t_{\pm}
	={}&
		\nBeff + \nbrack[\big]{1-\anis \cossq{\theta}}\nbrack[\big]{1-\nexJF_{\pm}\opbrack{\vk}}
	\\
	p_{\pm}
	={}&
		-\anis \sinsq{\theta} \nexJF_{\pm}\opbrack{\vk}
	\\
	\alpha
	={}&
		\frac{\nexJF_{\Re}\opbrack{\vk}}{\abs{\nexJF_{\Odd}\opbrack{\vk}}} \\
	\nBeff
	={}&
		\nBmag^{\zmcs} + \anis \nbrack{3\cossq{\theta}-1},
\end{align}
and
\begin{equation}
	\nexJF_{\pm}\opbrack{\vk}
	=
		\nexJF_{\Even}\opbrack{\vk} \mp \abs{\nexJF_{\Odd}\opbrack{\vk}}.
\end{equation}

The integrand to calculate the Curie temperature in \Eq{eq:CurieTempGen} reduces to
\begin{equation}
	\phi_{\Curie}
	=
		\frac{
			1+\anis
			+\nbrack{1-\anis}\nbrack[\big]{\nexJF_{\Re}\opbrack{\vk}-\nexJF_{\Even}\opbrack{\vk}}
		}{
			\nbrack[\big]{
				1+\anis - \nbrack{1-\anis} \nexJF_{\Even}\opbrack{\vk}
			}^{2}
			- \nbrack{1-\anis}^{2}
				\abs{\nexJF_{\Odd}\opbrack{\vk}}^{2}
		},
\end{equation}
in this case.

\section{Hexagonal lattice}
\label{SI::hex}

An hexagonal lattice can be considered as being a honeycomb lattice with only second nearest neighbor interactions ($\exJ\equiv\exJ_2, \exJ_1=\exJ_3=0$).
The integrand \eqref{eq:phi} then reduces to
\begin{equation} \label{eq:SIphiHex}
	\phi
	=
		\Mag \exJF\opbrack{0} \frac{t}{ 2 \omega }\cothfn[\Big]{ \frac{\omega}{2 \kBT}  } -\Over{2}.
\end{equation}
The quasi-particle excitation becomes
\begin{equation}
	\omega
	=
		\Mag \exJF\opbrack{0} \sqrt{t^2-p^2}.
\end{equation}
We introduced (with $\anis\coloneqq\anis_2$ and $\nexJF\opbrack{k}\coloneqq \nexJF_{\Even}\opbrack{\vk}$)
\begin{align}
	t
	\coloneqq{}&
		\nBeff + \nbrack{1-\anis \cossq{\theta}}\nbrack{1-\nexJF\opbrack{\vk}}
	\\
	p
	\coloneqq{}&
		-\anis \sinsq{\theta} \nexJF\opbrack{\vk} \\
	\nBeff
	\coloneqq{}&
		\nBmag^{\zmcs} + \anis \nbrack{3\cossq{\theta}-1}.
\end{align}
This, correctly, corresponds to the results obtained in Vanherck \andothers\citeSI{Vanherck18SI},\ with $S=\inOver{2}$ and with the normalized hexagonal Fourier transform $\nexJF\opbrack{\vk}$ replaced by its cubical equivalent.
The integrand to calculate the Curie temperature in \Eq{eq:CurieTempGen} reduces to
\begin{equation}
	\phi_{\Curie}
	=
		\Over{
			1+\anis
			-\nbrack{1-\anis} \nexJF\opbrack{\vk}
		},
\end{equation}
in this case.

\bibliographystyleSI{aipnum4-1.bst}
\bibliographySI{paper_apl_SI}

\end{document}